\documentclass[12pt]{article}
\usepackage{epsfig}
\newcommand{\mysection}{\setcounter{equation}{0}\section}

\def\beq{\begin{equation}}
\def\eeq{\end{equation}}
\def\beqa{\begin{eqnarray}}
\def\eeqa{\end{eqnarray}}

\newlength{\dinwidth} \newlength{\dinmargin}
\begin{document}

\begin{center}
{\Large \bf Charged Higgs production via $bg \longrightarrow t H^-$
at the LHC}
\end{center}
\vspace{2mm}
\begin{center}
{\large Nikolaos Kidonakis}\\
\vspace{2mm}
{\it Kennesaw State University\\
1000 Chastain Rd., \#1202, Kennesaw, GA 30144-5591}\\
\end{center}

\begin{abstract}
I present a calculation of  QCD radiative corrections to charged 
Higgs production via the process $bg \longrightarrow t H^-$.
I show that the cross section is dominated by soft-gluon corrections, 
which are computed through next-to-next-to-leading order.
Results for charged Higgs production at the LHC are presented,
including the dependence of the cross section on the charged Higgs mass,
the top quark mass, the factorization and renormalization scales,
and $\tan \beta$.

\end{abstract}

\mysection{Introduction}

One of the main goals of the current particle physics program
is the discovery of the Higgs boson.
The Minimal Supersymmetric Standard Model (MSSM) introduces charged Higgs 
bosons in addition to the neutral Higgs. A future discovery of a charged Higgs
would thus be a sure sign of new physics beyond the Standard Model.

In the MSSM there are two Higgs doublets, one giving mass to the 
up-type fermions and the other to the down-type fermions.
The ratio of the vacuum expectation values, $v_2,v_1$ for the two doublets
is $\tan \beta=v_2/v_1$.
Among the extra Higgs particles in the MSSM are two charged Higgs bosons,
$H^+$ and $H^-$. 

The Large Hadron Collider (LHC) at CERN has a good potential
for discovery of a charged Higgs boson. A promising channel
is associated production with a top quark via bottom-gluon fusion, 
$bg \longrightarrow t H^-$ [1-17]. 
In this paper we focus on $H^-$ production, but we note that
the cross sections for $H^+$ production, via the related process
${\bar b}g \longrightarrow {\bar t} H^+$, are identical.
The complete next-to-leading order (NLO) QCD corrections to the process 
$bg \longrightarrow t H^-$ have been recently derived in
Refs. \cite{Zhu,Plehn,BHJP}. These corrections were shown to
stabilize the cross section with respect to changes in factorization
and renormalization scales. The SUSY-QCD NLO corrections were also
calculated in \cite{Plehn,BHJP}.

The NLO QCD corrections were shown to be substantial, up to
85\% enhancement of the lowest order cross section \cite{Zhu}.
The SUSY-QCD corrections are comparatively small, though
non-negligible, and their precise value
depends on several parameters of the MSSM \cite{Plehn}.
Since the NLO QCD corrections are large it is important to 
consider whether even higher-order corrections may make a significant
contribution. In this paper I show that the NLO corrections
are dominated by near-threshold soft-gluon emission and I calculate
the contribution from next-to-next-to-leading order (NNLO)
soft-gluon corrections, which are sizable.

The charged Higgs is expected to be quite massive, so its production
at current colliders will be a near-threshold process. In such processes
soft gluon emission is expected to dominate the radiative corrections.
This has in fact been shown by now for a large number of processes
including top, bottom, and charm quark production \cite{NKtop},
$W$-boson production \cite{NKASV}, direct photon production \cite{NKJOph},
jet production \cite{NKJOjet}, and flavor-changing-neutral-current 
single-top production \cite{NKAB}.
Near threshold for the production of a specified final state
there is limited energy available for the production
of any additional radiation; hence the emitted gluons are soft and they 
manifest themselves in logarithmically enhanced terms that numerically
dominate the cross section. 
The structure of these threshold contributions follows from general
considerations of the factorization properties of hard-scattering 
cross sections \cite{KS,LOS,NKuni}. 
Renormalization properties of the factorized pieces of
a cross section result in formal resummations, which provide the
form of the soft-gluon corrections to all orders in the 
strong coupling $\alpha_s$. 
For further details and reviews see Refs. \cite{KS,NKuni,NKMPLA}.

\mysection{NNLO soft-gluon corrections}

In this section we derive the analytical form of the 
soft-gluon corrections through next-to-next-to-leading order  
for charged Higgs production with a top quark
in hadronic collisions.
For the process $b(p_b)+g(p_g) \rightarrow t(p_t)+H^-(p_{H^-})$,
we define the kinematical invariants $s=(p_b+p_g)^2$,
$t=(p_b-p_t)^2$, $u=(p_g-p_t)^2$, and $s_4=s+t+u-m_t^2-m^2_{H^-}$,
where $m_{H^-}$ is the charged Higgs mass and $m_t$ is the top quark mass. 
Note that we ignore the mass of the $b$-quark in the kinematics.
Near threshold, i.e. when we have just enough
partonic energy to produce the $tH^-$ final state,  $s_4 \rightarrow 0$.
The threshold corrections then take the form of logarithmic plus 
distributions, $[(\ln^l(s_4/m_{H^-}^2)/s_4)]_+$, 
where $l\le 2n-1$ for the $n$-th order QCD corrections. 
These plus distributions are defined 
through their integral with any smooth function, such as parton distributions, 
giving a finite result.
The leading logarithms (LL) are those with $l=2n-1$ while the
next-to-leading logarithms (NLL) are those with $l=2n-2$.
In this paper we calculate NLO and NNLO soft-gluon threshold corrections
at NLL accuracy, i.e. at each order including both leading and next-to-leading
logarithms. We denote them as NLO-NLL and NNLO-NLL, respectively.  
Thus, at NLO we include $[\ln(s_4/{m^2_{H^-}})/s_4]_+$ (LL)
and $[1/s_4]_+$ (NLL) terms. Although we do not calculate the full virtual
$\delta(s_4)$ terms, we include those $\delta(s_4)$ terms that involve the
factorization and renormalization scales, denoted by $\mu_F$ and
$\mu_R$ respectively.
At NNLO, we include $[\ln^3(s_4/{m^2_{H^-}})/s_4]_+$ (LL)
and $[\ln^2(s_4/{m^2_{H^-}})/s_4]_+$ (NLL) terms. We also include
some $[\ln(s_4/{m^2_{H^-}})/s_4]_+$ and $[1/s_4]_+$ terms
that involve the factorization and renormalization scales; and some constants
which arise from the inversion from moment space, where the resummation
is performed, back to momentum space. For details of this approach see
Refs. \cite{NKtop,NKuni}.

The differential Born cross section is 
$d^2{\hat\sigma}^B_{bg \rightarrow t H^-}/(dt \; du)
=F^B_{bg \rightarrow t H^-} \delta(s_4)$
where 
\beqa
F^B_{bg \rightarrow t H^-}&=&
\frac{\pi \alpha \alpha_s (m_b^2 \tan^2\beta
+m_t^2 \cot^2\beta)}{12 s^2 m_W^2 \sin^2\theta_W}
\left\{\frac{s+t-m^2_{H^-}}{2s} \right. 
\nonumber \\ && \hspace{-15mm} \left.
{}-\frac{m_t^2(u-m^2_{H^-})+m^2_{H^-}(t-m_t^2)+s(u-m_t^2)}{s(u-m_t^2)}
-\frac{m_t^2(u-m^2_{H^-}-s/2)+su/2}{(u-m_t^2)^2}\right\} ,
\label{FBorn}
\eeqa
where $\alpha=e^2/(4\pi)$, $\alpha_s$ is the strong coupling, 
and we have kept the $b$-quark mass, $m_b$, 
non-zero only in the $m_b^2 \tan^2 \beta$ term. 
We use consistently the running masses for the top and bottom
quarks \cite{DKS}, corresponding to pole masses of 175 GeV and 4.8 GeV, 
respectively.

We next proceed with the calculation of the NLO and NNLO soft-gluon
corrections at NLL accuracy. In our derivation of these corrections
we follow the general techniques and master formulas presented
in Ref. \cite{NKuni}.

The NLO soft-gluon corrections for the process $bg \rightarrow tH^-$ are
\beq
\frac{d^2{\hat\sigma}^{(1)}_{bg\rightarrow t H^-}}{dt \, du}
=F^B_{bg \rightarrow t H^-}
\frac{\alpha_s(\mu_R^2)}{\pi} \left\{
c^{bg \rightarrow tH^-}_{3} \left[\frac{\ln(s_4/m^2_{H^-})}{s_4}\right]_+
+c^{bg \rightarrow tH^-}_{2} \left[\frac{1}{s_4}\right]_+
+c^{bg \rightarrow tH^-}_{1}  \delta(s_4)\right\} \, .
\label{NLObgtH}
\eeq

Here $c^{bg \rightarrow tH^-}_{3}=2(C_F+C_A)$, where $C_F=(N_c^2-1)/(2N_c)$
and $C_A=N_c$ with $N_c=3$ the number of colors,  and
\beqa
c^{bg \rightarrow tH^-}_{2}&=&2 {\rm Re} {\Gamma'}_S^{(1)}
-C_F-C_A-2C_F\ln\left(\frac{-u+m^2_{H^-}}{m^2_{H^-}}\right)
-2C_A\ln\left(\frac{-t+m^2_{H^-}}{m^2_{H^-}}\right)
\nonumber \\ &&
{}-(C_F+C_A)\ln\left(\frac{\mu_F^2}{s}\right)
\nonumber \\
&\equiv& T^{bg \rightarrow tH^-}_{2}-(C_F+C_A)
\ln\left(\frac{\mu_F^2}{m^2_{H^-}}\right) \, ,
\eeqa
where $\mu_F$ is the factorization scale, and we have defined 
$T^{bg \rightarrow tH^-}_{2}$ as the scale-independent part
of $c^{bg \rightarrow tH^-}_{2}$.
The term ${\rm Re} {\Gamma'}_S^{(1)}$ denotes the real part
of the one-loop soft anomalous dimension, which describes
noncollinear soft-gluon emission \cite{KS}, modulo some gauge-dependent
terms that cancel out in the cross section. A one-loop 
calculation gives 
\beq
{\Gamma'}_S^{(1)}=C_F \ln\left(\frac{-t+m_t^2}{m_t\sqrt{s}}\right)
+\frac{C_A}{2} \ln\left(\frac{-u+m_t^2}{-t+m_t^2}\right)
+\frac{C_A}{2} (1-\pi i).
\eeq
Also
\beqa
c^{bg \rightarrow tH^-}_{1}&=&\left[C_F \ln\left(\frac{-u+m^2_{H^-}}
{m^2_{H^-}}\right)
+C_A \ln\left(\frac{-t+m^2_{H^-}}{m^2_{H^-}}\right)
-\frac{3}{4}C_F-\frac{\beta_0}{4}\right]
\ln\left(\frac{\mu_F^2}{m^2_{H^-}}\right)
\nonumber \\ &&
{}+\frac{\beta_0}{4} \ln\left(\frac{\mu_R^2}{m^2_{H^-}}\right) \, ,
\eeqa
where $\mu_R$ is the renormalization scale and $\beta_0=(11C_A-2n_f)/3$ 
is the lowest-order $\beta$ function, 
with $n_f$ the number of light quark flavors.
Note that $c^{bg \rightarrow tH^-}_{1}$ represents the scale-dependent
part of the $\delta(s_4)$ corrections. We do not calculate the full virtual
corrections here. Our calculation of the NLO soft-gluon corrections
includes the full leading and next-to-leading logarithms (NLL) and is thus
a NLO-NLL calculation. 

We next calculate the NNLO soft-gluon corrections for $bg \rightarrow tH^-$:
\beqa
&& \hspace{-5mm}\frac{d^2{\hat\sigma}^{(2)}_{bg \rightarrow tH^-}}
{dt \, du}
=F^B_{bg \rightarrow tH^-} \frac{\alpha_s^2(\mu_R^2)}{\pi^2} 
\left\{\frac{1}{2} \left(c^{bg \rightarrow tH^-}_{3}\right)^2 
\left[\frac{\ln^3(s_4/m^2_{H^-})}{s_4}\right]_+ \right.
\nonumber \\ && \hspace{-5mm}
{}+\left[\frac{3}{2} c^{bg \rightarrow tH^-}_{3} \, c^{bg 
\rightarrow tH^-}_{2}
-\frac{\beta_0}{4} c^{bg \rightarrow tH^-}_{3} \right] 
\left[\frac{\ln^2(s_4/m^2_{H^-})}{s_4}\right]_+ 
\nonumber \\ && \hspace{-5mm}
{}+\left[c^{bg \rightarrow tH^-}_{3} \, c^{bg \rightarrow tH^-}_{1}
+(C_F+C_A)^2\ln^2\left(\frac{\mu_F^2}{m^2_{H^-}}\right)
-2(C_F+C_A) T_2^{bg \rightarrow tH^-}\ln\left(\frac{\mu_F^2}{m^2_{H^-}}\right)
\right.
\nonumber \\ && \quad \left.
{}+\frac{\beta_0}{4} c^{bg \rightarrow tH^-}_{3} 
\ln\left(\frac{\mu_R^2}{m^2_{H^-}}\right)
-\zeta_2 \, \left(c^{bg \rightarrow tH^-}_{3}\right)^2 \right]
\left[\frac{\ln(s_4/m^2_{H^-})}{s_4}\right]_+
\nonumber \\ && \hspace{-5mm} 
{}+\left[-(C_F+C_A) \ln\left(\frac{\mu_F^2}{m^2_{H^-}}\right)
c^{bg \rightarrow tH^-}_{1}
-\frac{\beta_0}{4} (C_F+C_A) \ln\left(\frac{\mu_F^2}{m^2_{H^-}}\right) 
\ln\left(\frac{\mu_R^2}{m^2_{H^-}}\right) \right.
\nonumber \\ && \quad \left. \left.
{}+(C_F+C_A)\frac{\beta_0}{8} \ln^2\left(\frac{\mu_F^2}{m^2_{H^-}}\right)
-\zeta_2 \, c^{bg \rightarrow tH^-}_{2} \, c^{bg \rightarrow tH^-}_{3}
+\zeta_3 \, \left(c^{bg \rightarrow tH^-}_{3}\right)^2\right]
\left[\frac{1}{s_4}\right]_+ \right\} \, ,
\label{NNLObgtH}
\eeqa
where $\zeta_2=\pi^2/6$ and $\zeta_3=1.2020569...$.
We note that only the leading and next-to-leading logarithms are complete.
Hence this is a NNLO-NLL calculation.
Consistent with a NLL calculation we have also kept all logarithms of the 
factorization and renormalization scales in the
$[\ln(s_4/m^2_{H^-})/s_4]_+$ terms, and squares of logarithms
involving the scales in the $[1/s_4]_+$ terms, 
as well as $\zeta_2$ and $\zeta_3$ terms
that arise in the calculation of the soft corrections 
when inverting from moments back to momentum space \cite{NKtop,NKuni}.

In principle one can obtain the form of the soft radiative corrections
at any order in $\alpha_s$ and indeed resum them to all orders.
However in practice such resummed cross sections depend on a prescription
to avoid the infrared singularity and ambiguities from prescription
dependence can actually be larger than contributions from terms
beyond NNLO \cite{NKtop}. Hence we here give results to NNLO as has been done
for many other processes \cite{NKMPLA,NKtop,NKASV,NKJOph,NKJOjet,NKAB,NKuni}.

We now convolute the partonic cross sections with parton distribution
functions to obtain the hadronic cross section in $pp$ collisions at the LHC.
For the hadronic cross section $p(p_A)+p(p_B) 
\rightarrow t(p_t)+H^-(p_{H^-})$ we define 
$S=(p_A+p_B)^2$, $T=(p_A-p_t)^2$, and $U=(p_B-p_t)^2$,
and note that $p_b=x_A p_A$, $p_g=x_B p_B$, where $x$ denotes
the momentum fraction of the hadron carried by the parton.
The hadronic cross section is then given by
\beqa
\sigma_{pp \rightarrow tH^-}(S)&=&
\int_{T_{min}}^{T_{max}} dT 
\int_{-S-T+m_t^2+m^2_{H^-}}^{m_t^2+m_t^2 S/(T-m_t^2)} dU 
\int_{(m^2_{H^-}-T)/(S+U-m_t^2)}^1 dx_B \int_0^{x_B(S+U-m_t^2)+T-m^2_{H^-}} 
ds_4
\nonumber \\ &&
\times \frac{x_A x_B}{x_B S+T-m_t^2} \,
\phi(x_A) \, \phi(x_B) \, 
\frac{d^2{\hat\sigma}_{bg \rightarrow tH^-}}{dt \, du}
\eeqa
where
\beq
x_A=\frac{s_4-m_t^2+m^2_{H^-}-x_B(U-m_t^2)}{x_B S+T-m_t^2},
\eeq
$T_{^{max}_{min}}=-(1/2)(S-m_t^2-m^2_{H^-}) \pm 
(1/2) \sqrt{(S+m_t^2-m^2_{H^-})^2-4m_t^2S}$, and
$\phi(x)$ are the parton distributions.

\mysection{Charged Higgs production at the LHC}

\begin{figure}
\begin{center}
\includegraphics[width=12.5cm]{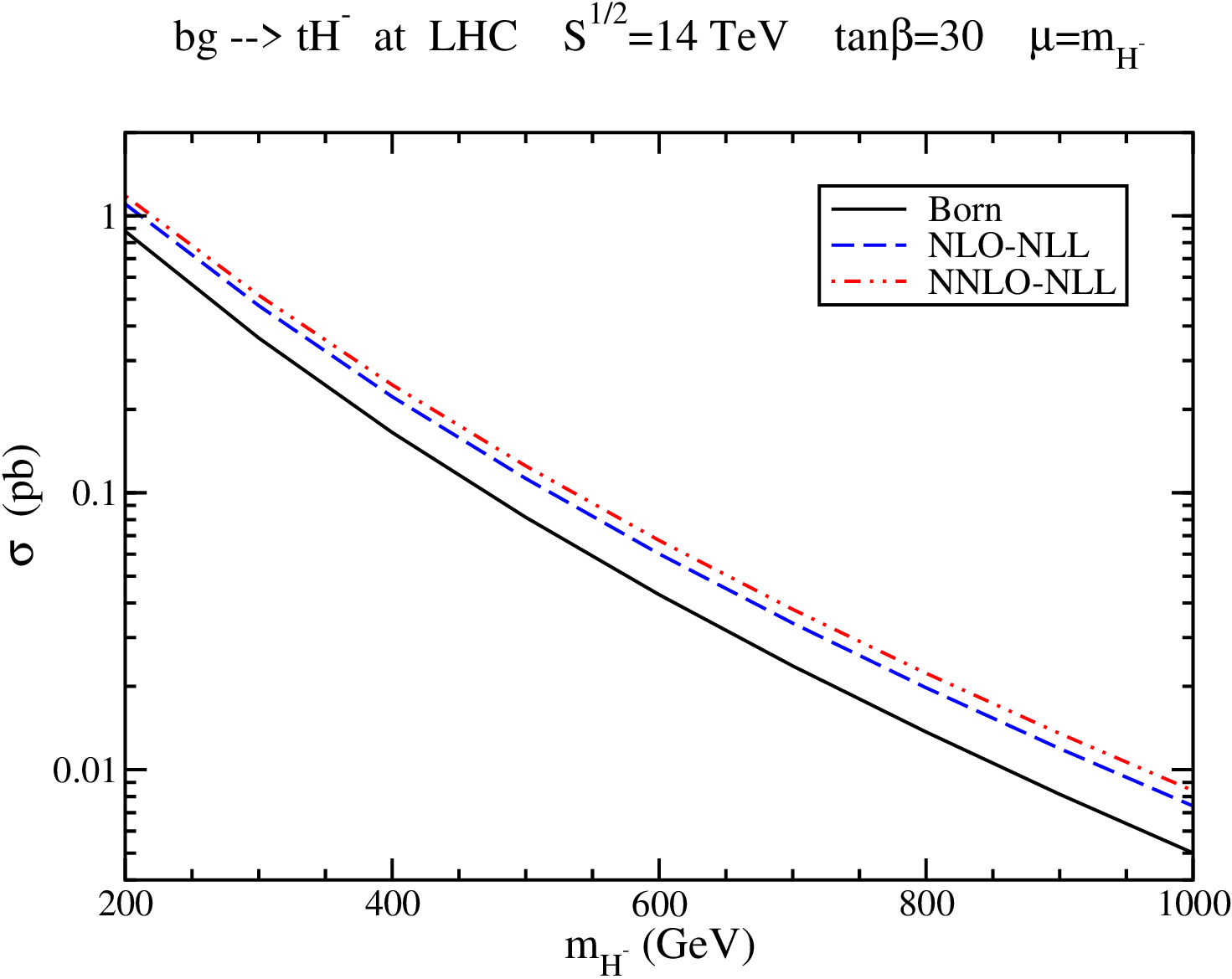} 
\caption{The total cross section for charged Higgs production at the LHC.}
\label{higgsmhplot}
\end{center}
\end{figure}

We now turn our attention to detailed numerical results for charged
Higgs production at the LHC.
In Figure \ref{higgsmhplot} we plot the cross section versus
charged Higgs mass for $pp$ collisions at the LHC with $\sqrt{S}=14$ TeV. 
Here and throughout this paper we use the MRST2002 approximate 
NNLO parton distributions \cite{mrst2002}
with the respective  three-loop evaluation of $\alpha_s$. 
We set the factorization scale equal to the renormalization scale 
and denote this common scale by $\mu$.
We show results for the Born, NLO-NLL, and NNLO-NLL
cross sections, all with a choice of scale $\mu=m_{H^-}$. 
We use the same NNLO parton densities and couplings in all the results,
so that we can concentrate on the effects of the soft-gluon corrections.
We choose a value $\tan \beta=30$. 
It is straightforward to get the results for any other value of
$\tan \beta$, since the only dependence on $\beta$ in our equations
is in the factor $m_b^2 \tan^2 \beta+m_t^2 \cot^2 \beta$ appearing
in the Born term, Eq. \ref{FBorn}.
The cross sections span over two orders of magnitude in the mass
range shown, 200 GeV $\le m_{H^-} \le$ 1000 GeV.
The NLO and NNLO threshold corrections are positive and provide a significant
enhancement to the lowest-order result.
We note that the cross sections for the related process
${\bar b} g \rightarrow {\bar t} H^+$ are exactly the same.  

\begin{figure}
\begin{center}
\includegraphics[width=12.5cm]{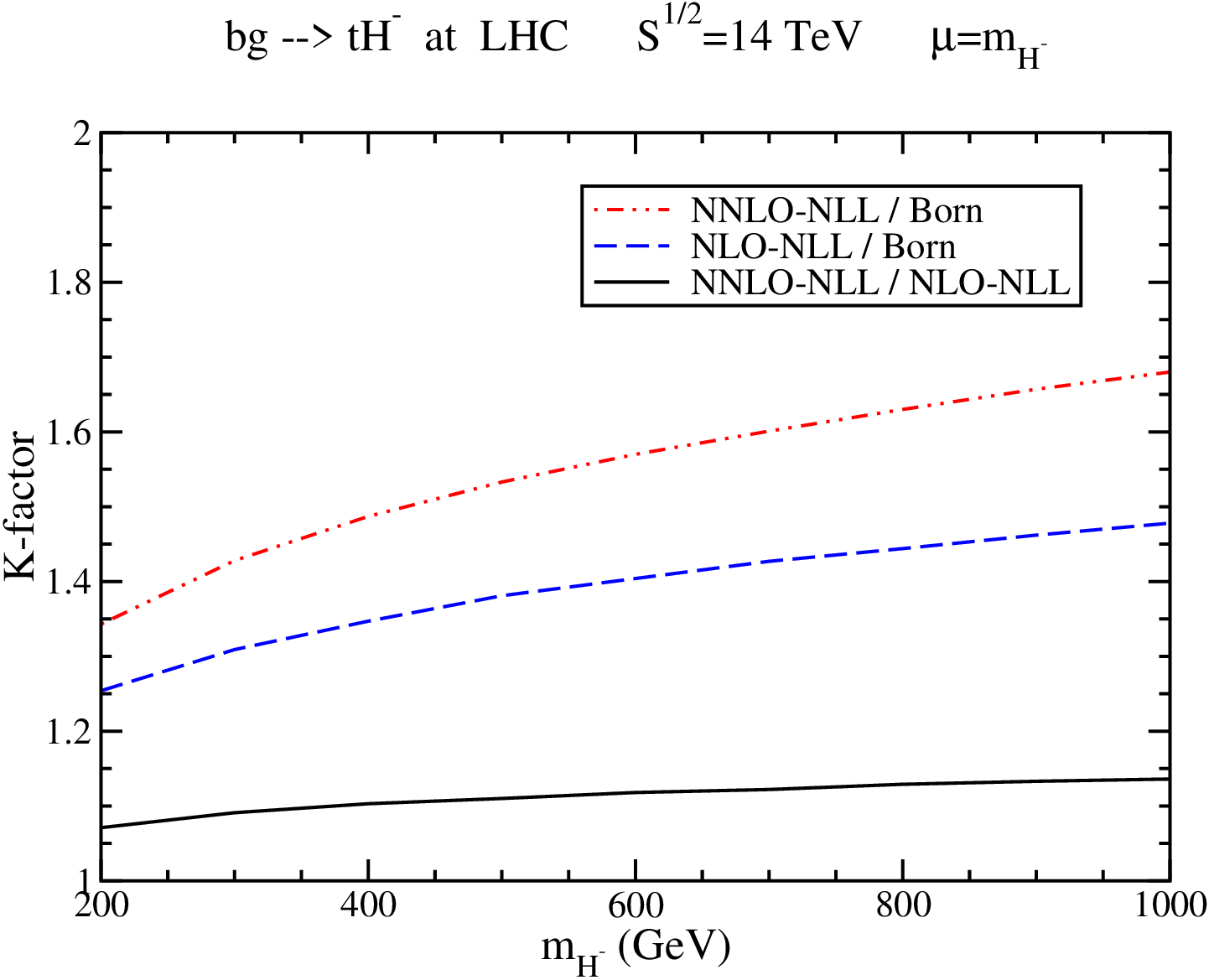} 
\caption{The $K$-factors for charged Higgs production at the LHC.}
\label{Khiggsmhplot}
\end{center}
\end{figure}

The relative size of the corrections is better shown in Figure 
\ref{Khiggsmhplot} where we plot the $K$-factors, i.e. ratios of cross
sections at various orders. The NLO-NLL / Born curve shows that the
NLO threshold corrections enhance the Born cross section by approximately
25\% to 50\% depending on the mass of the charged Higgs. As expected
the corrections increase for higher charged Higgs masses since then 
we get closer to threshold.
The NNLO-NLL / Born curve shows that if we include the NNLO threshold 
corrections we get an enhancement over the Born result of approximately
35\% to 70\% in the range of masses shown. Again the enhancement increases
with charged Higgs mass, as expected.
Finally, the NNLO-NLL / NLO-NLL curve shows clearly the further enhancement 
over NLO that the NNLO threshold corrections provide. This curve
is simply the ratio of the other two curves and varies between 7\% and
14\%.

We now want to compare our NLO-NLL results with the exact results
that have been derived in \cite{Zhu,Plehn}. We note that different
choices of factorization/renormalization scales were used in those
references. In Ref. \cite{Zhu} the reference scale chosen was
$m_{H^-}+m_t$ while in Ref. \cite{Plehn} it was $(m_{H^-}+m_t)/2$. 
In this paper we choose $m_{H^-}$. This is the natural choice in our
approach since we are considering logarithms of $s_4/m^2_{H^-}$.
Of course any choice of scale is theoretically possible
and a cross section known to all orders does not depend on the scale.
However a finite-order cross section does depend on the scale, though the
dependence decreases as we move from Born to NLO to NNLO and so on.
The work in \cite{Zhu,Plehn} indeed showed a reduction of scale 
dependence when the NLO corrections are added relative to the Born
cross section.
In fact, as we will see below, the NNLO threshold corrections further decrease
the scale dependence, thus resulting in more stable predictions.

\begin{figure}
\begin{center}
\includegraphics[width=12cm]{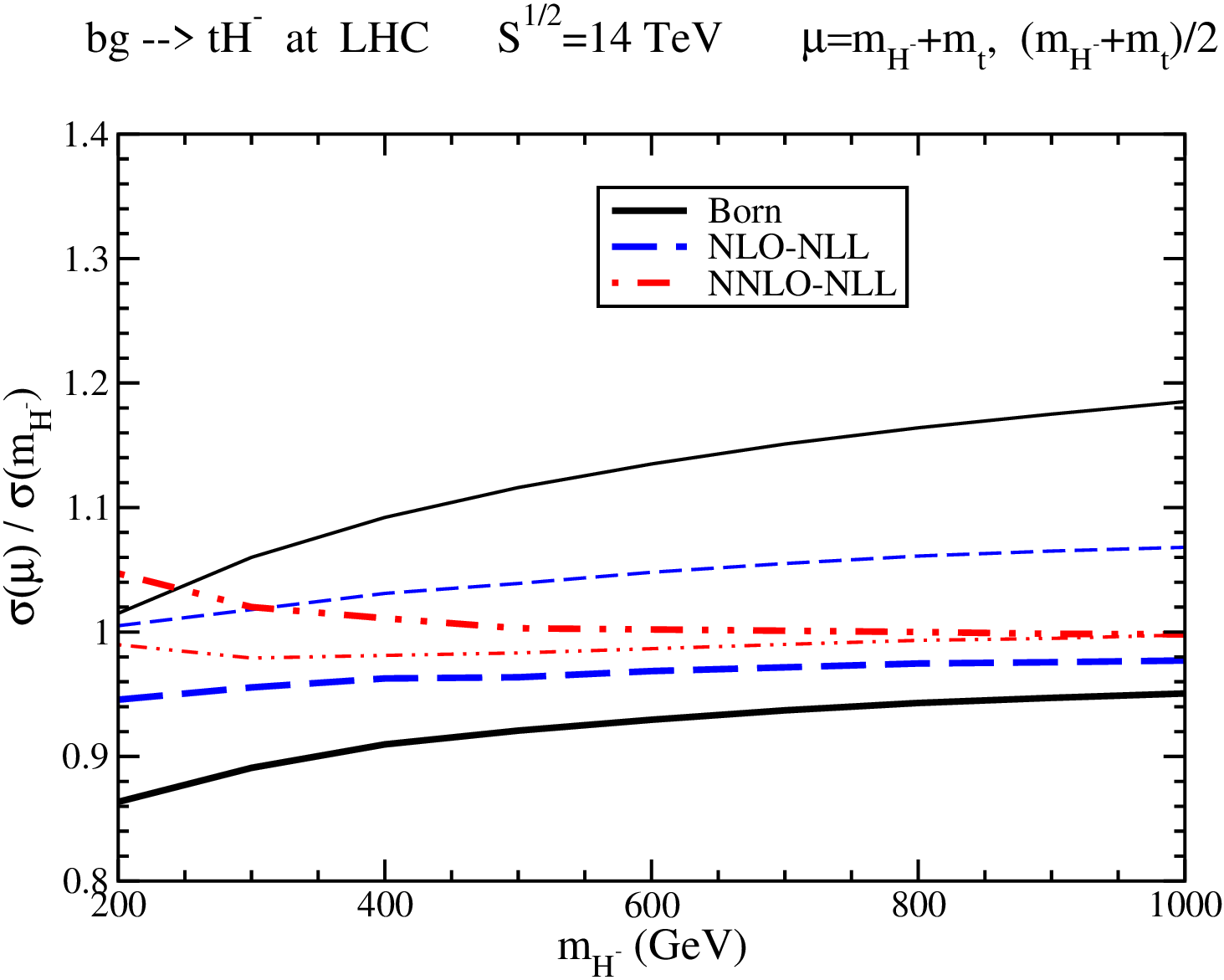} 
\caption{The ratio of cross sections
at various scales for charged Higgs production at the LHC. 
The bold lines are with $\mu=m_{H^-}+m_t$; the rest are with 
$\mu=(m_{H^-}+m_t)/2$.}
\label{higgsratmhmtovermhplot}
\end{center}
\end{figure}

Before comparing our results to the exact NLO cross section, we can check
the effect of choosing the scales used in Refs. \cite{Zhu,Plehn}. 
In Figure \ref{higgsratmhmtovermhplot} we plot the ratios of the 
cross sections with choice of reference scale $\mu=m_{H^-}+m_t$
(bold lines) and $\mu=(m_{H^-}+m_t)/2$ (thin lines) over
the cross section with scale $\mu=m_{H^-}$.
We see that indeed there is a considerable variation at lowest order,
but this progressively diminishes at NLO and NNLO.
In fact at NNLO there is hardly any difference between the two lines
at large values of the charged Higgs mass.
Thus we see the stabilization of the cross section versus scale variation
when higher-order corrections are included. We will say more regarding 
this important point and show more plots below.

\begin{figure}
\begin{center}
\includegraphics[width=12cm]{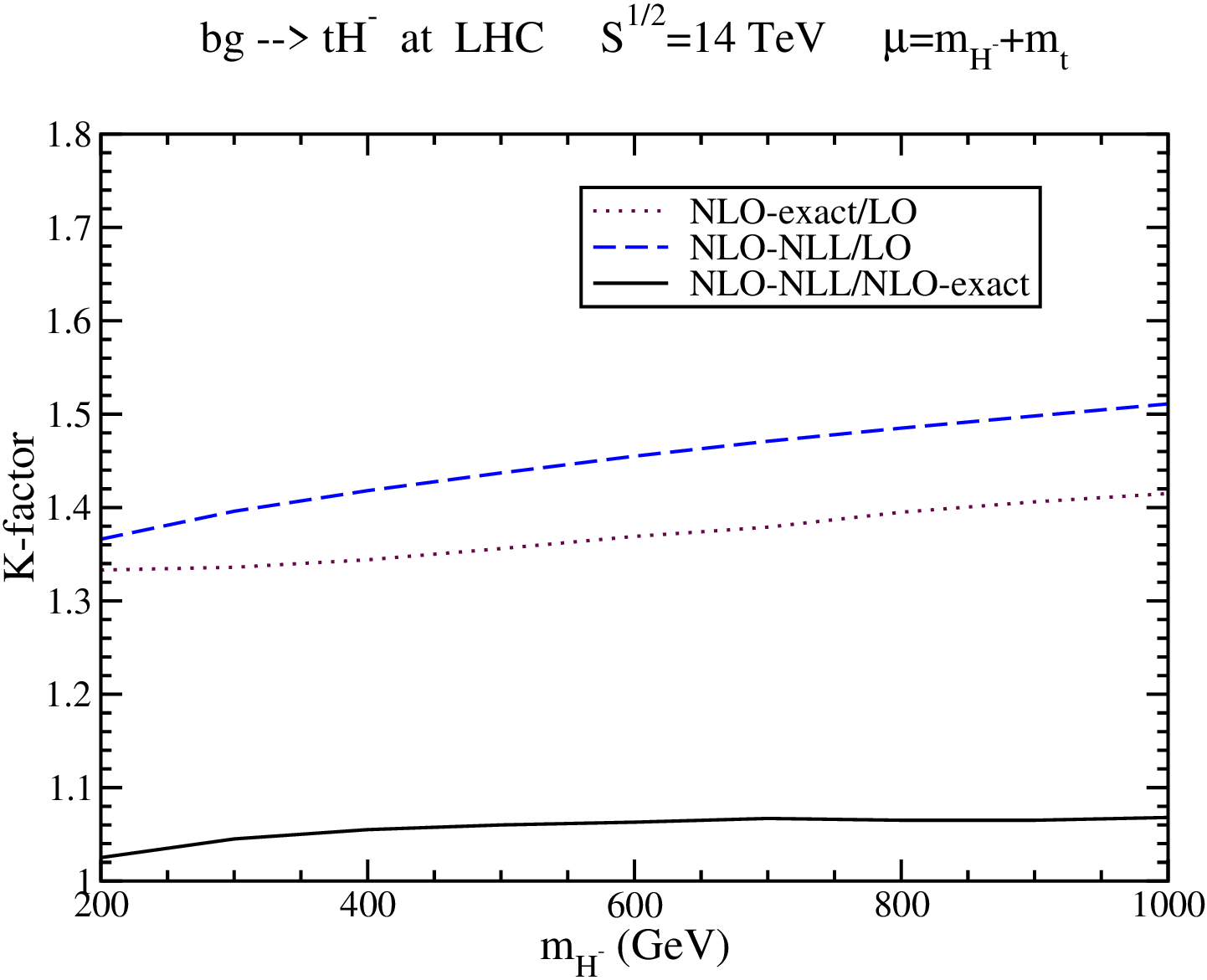} 
\caption{The ratio of exact and approximate NLO cross sections 
for charged Higgs production at the LHC.}
\label{higgsNLOcomparplot}
\end{center}
\end{figure}

We now compare the NLO soft-gluon results with the exact NLO cross section.
In Figure \ref{higgsNLOcomparplot}, we compare the NLO-NLL cross section
with the exact NLO cross section of reference \cite{Zhu}. To make the 
comparison, the NLO-NLL result is calculated here for $\mu=m_{H^-}+m_t$
since that's the scale chosen in \cite{Zhu} and also using a two-loop
$\alpha_s$. Also, to remove discrepancies
arising from different choices of parton distribution functions,
we plot $K$-factors. 
The NLO-exact / LO curve is taken from Ref. \cite{Zhu} by dividing
curve 1 by curve 2 in Figure 6 of that reference (to account for the 
different definition of $K$-factor used there).
The fact that the NLO-NLL / NLO-exact curve 
is very close to 1 (only a few percent difference) shows 
that the NLO-NLL cross section is a 
remarkably good approximation to the exact NLO result. As noted before,
we might have expected this on theoretical grounds since this is 
near-threshold production, and also from prior experience with many
other near-threshold hard-scattering cross sections
\cite{NKtop,NKASV,NKJOph,NKJOjet,NKAB}.

\begin{figure}
\begin{center}
\includegraphics[width=12cm]{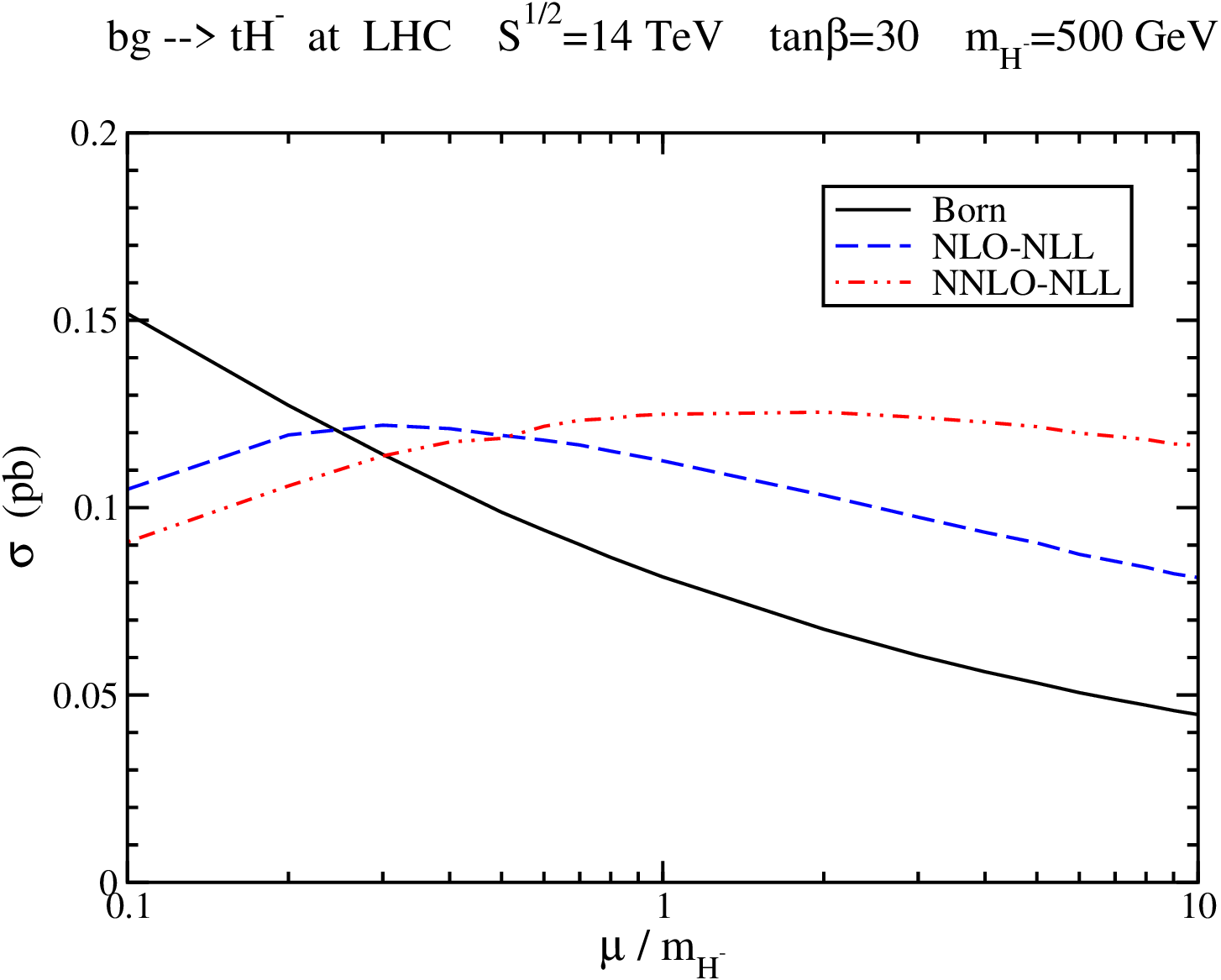} 
\caption{The scale dependence for production of a charged Higgs 
with mass $m_{H^-}=500$ GeV at the LHC.}
\label{higgsmuplot}
\end{center}
\end{figure}

In Figure \ref{higgsmuplot}, we plot the scale dependence of the
cross section for a fixed charged Higgs mass $m_{H^-}=500$ GeV
and $\tan \beta=30$. 
We plot a large range in scale, $0.1 \le \mu/m_{H^-} \le 10$,
and see indeed
that the threshold corrections greatly decrease the scale dependence 
of the cross section. The NNLO-NLL curve is relatively flat.

\begin{figure}
\begin{center}
\includegraphics[width=12cm]{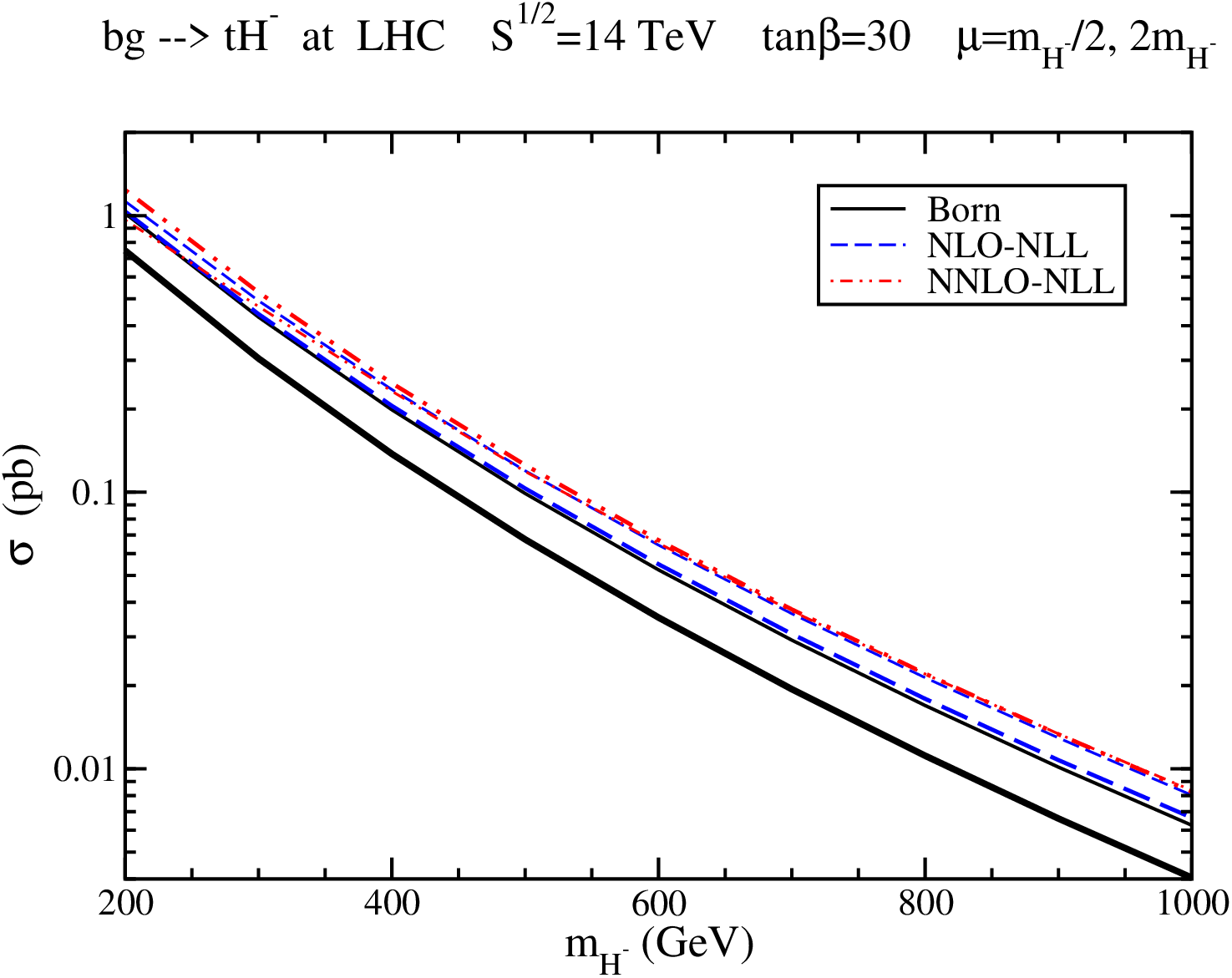} 
\caption{The scale dependence for charged Higgs production at the LHC.
The bold lines are with $\mu=2m_{H^-}$; the rest are with $\mu=m_{H^-}/2$.}
\label{higgsmhmuplot}
\end{center}
\end{figure}

In Figure \ref{higgsmhmuplot}  we plot the the cross section 
as a function of charged Higgs mass with two different
choices of scale, $\mu=m_{H^-}/2$ and $2m_{H^-}$.
We see that the variation
with scale of the Born cross section is quite large. The variation at
NLO-NLL is smaller, and at NNLO-NLL it is very small.
In fact the two NNLO-NLL curves are on top of each other for most
of the range in $m_{H^-}$.
Hence, the scale dependence of the cross section is drastically reduced when
higher-order corrections are included.
This is as expected from and is consistent with the reduced scale
dependence shown in Figures 
\ref{higgsratmhmtovermhplot} and \ref{higgsmuplot}.

\begin{figure}
\begin{center}
\includegraphics[width=12cm]{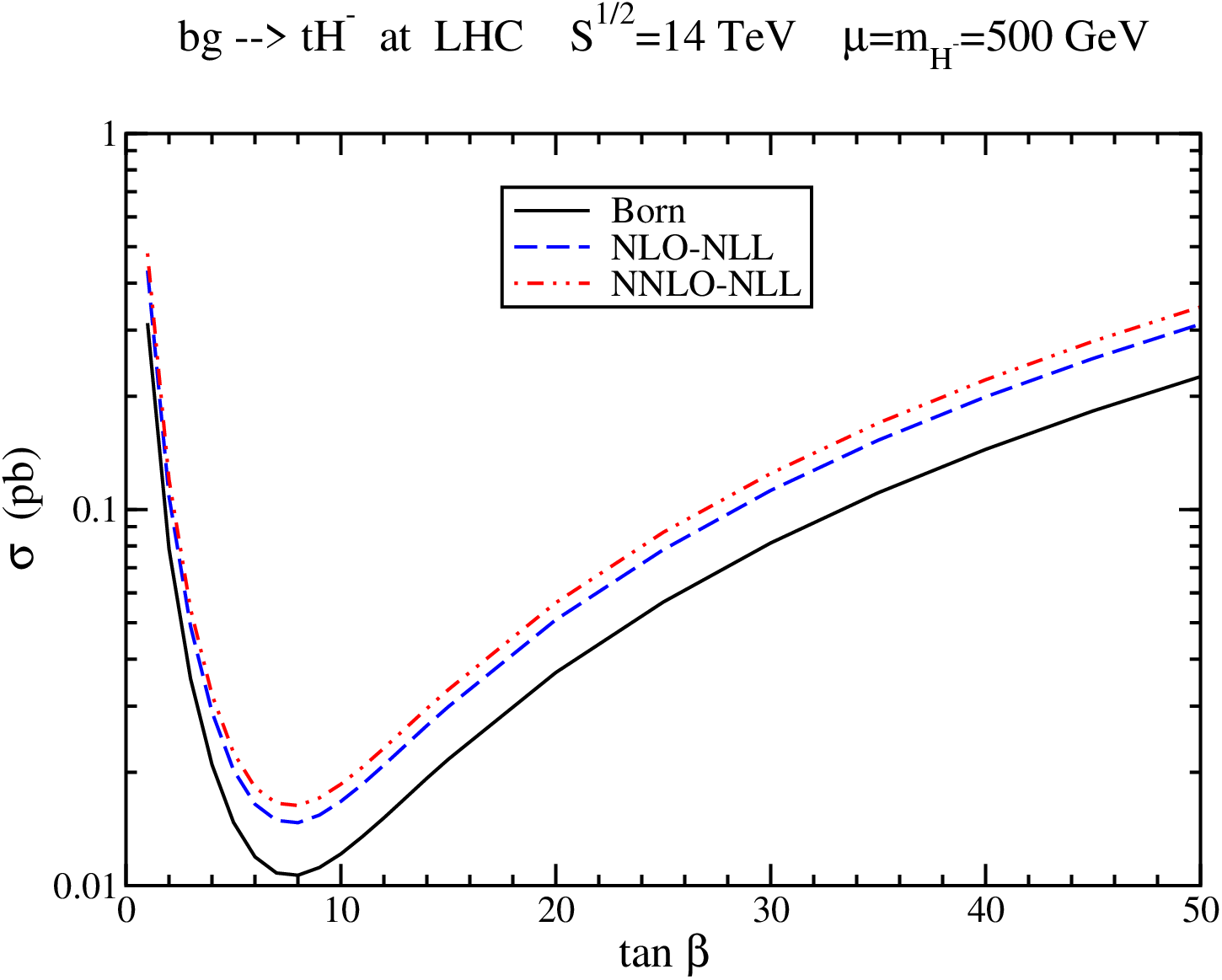} 
\caption{The $\tan \beta$ dependence for charged Higgs production at the LHC.}
\label{higgstanbplot}
\end{center}
\end{figure}

In Figure \ref{higgstanbplot} we plot the dependence of the cross
section on $\tan \beta$, over the range $1 \le \tan \beta \le 50$,
for fixed charged Higgs mass and scale 
$\mu=m_{H^-}=500$ GeV. The cross section is at a minimum near
$\tan\beta=8$. We note that the $\tan \beta$ dependence arises in the
factor $m_b^2 \tan^2 \beta + m_t^2 \cot^2 \beta$ in the Born term
and thus the shape of the curves at higher orders is similar. 
The dependence on $\tan \beta$ is quite large, spanning nearly two
orders of magnitude in the range shown.

\begin{figure}
\begin{center}
\includegraphics[width=12cm]{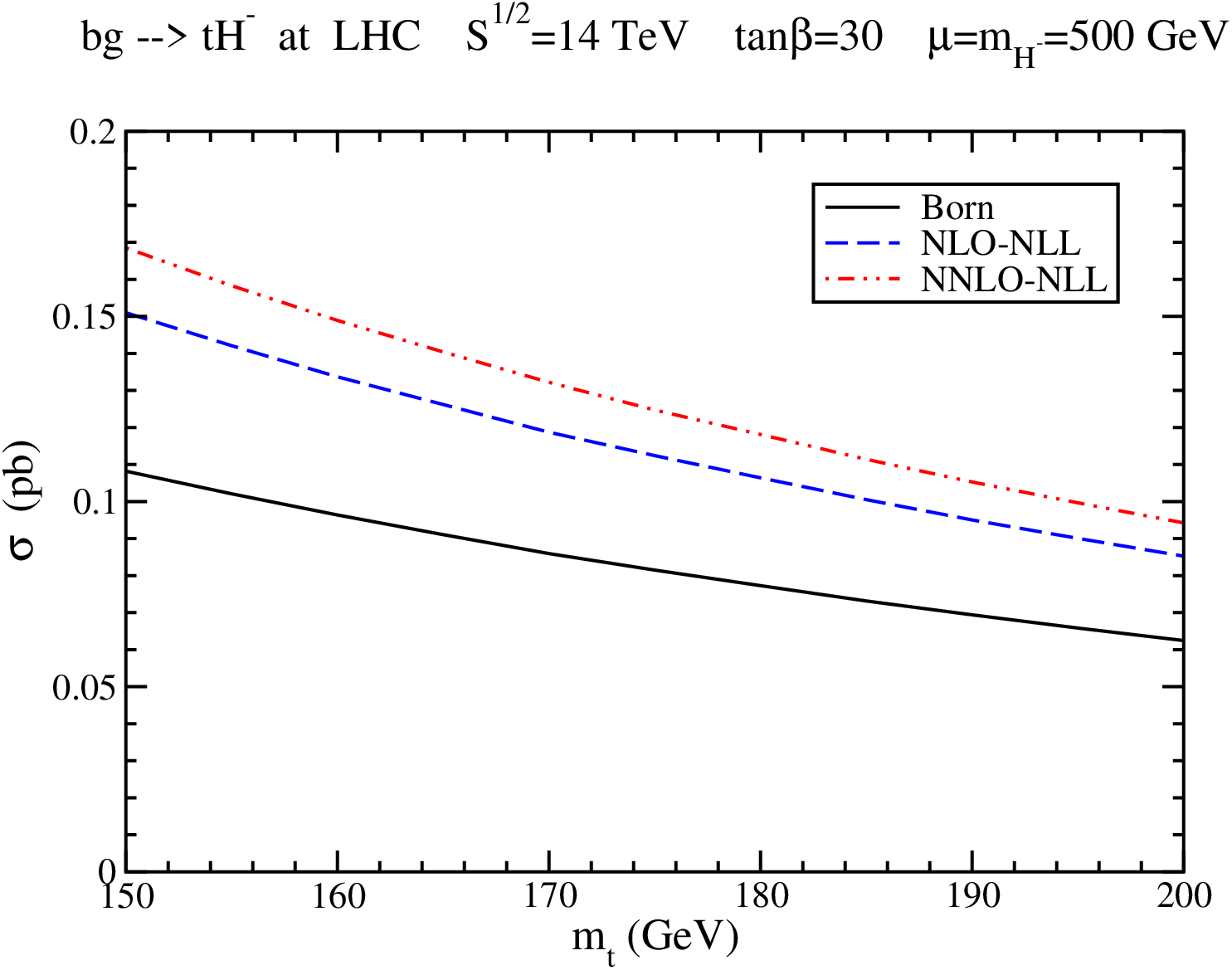} 
\caption{The top quark mass dependence for charged Higgs 
production at the LHC.}
\label{higgstopplot}
\end{center}
\end{figure}

In Figure \ref{higgstopplot} we plot the dependence of the cross section
on the top quark mass for fixed charged Higgs mass and scale 
$\mu=m_{H^-}=500$ GeV and $\tan \beta=30$. 
We see that the dependence is not very strong
so that the present experimental uncertainties on the top quark mass 
do not play a dominant role in the total uncertainty of the charged 
Higgs production cross section. As the top quark mass gets more precisely
known, this dependence will diminish further.

\mysection{Conclusion}

The process $bg \rightarrow t H^-$ offers a promising possibility for
discovering a charged Higgs boson.
Charged Higgs production at the LHC receives important contributions from
the threshold region. The NLO corrections to the process
$bg \rightarrow t H^-$ are quite large.
We have seen that the full NLO cross section is very 
well approximated by the NLO-NLL soft-gluon result, to within a few percent.
The NNLO soft-gluon threshold corrections to charged
Higgs production are important and further stabilize the cross section
versus changes in factorization and renormalization scales.
The dependence on $\tan \beta$ and on the charged Higgs mass
are quite large while the dependence on the top quark mass is milder.

\end{document}